\documentclass[journal]{IEEEtranTIE}
\usepackage{graphicx}
\usepackage{cite}
\usepackage{picinpar}
\usepackage{amsmath}
\usepackage{url}
\usepackage{flushend}
\usepackage[latin1]{inputenc}
\usepackage{colortbl}
\usepackage{soul}
\usepackage{multirow}
\usepackage{pifont}
\usepackage{color}
\usepackage{alltt}
\usepackage[hidelinks]{hyperref}
\usepackage{enumerate}
\usepackage{siunitx}
\usepackage{breakurl}
\usepackage{epstopdf}
\usepackage{pbox}

\usepackage{amssymb} 
\usepackage{booktabs}
\usepackage{url}
\usepackage{xcolor}
\usepackage{float}
\usepackage{optidef}
\usepackage{mathtools}
\usepackage{nomencl}

\usepackage{array}
\usepackage{caption}
\usepackage{subcaption}

\usepackage{amsthm}
\theoremstyle{remark}
\newtheorem{remark}{Remark}
\usepackage{multirow}
\usepackage[ruled,vlined]{algorithm2e}
\usepackage{hhline}
\usepackage{array}
\usepackage{multirow}
\usepackage{threeparttable}
\usepackage{acronym}
\usepackage{algpseudocode} 

\begin{document}
\title{Battery State of Health Estimation and Incremental Capacity Analysis under Dynamic Charging Profile Using Neural Networks}

\author{
	\vskip 1em
	
	Qinan Zhou,
	Gabrielle Vuylsteke,
    R. Dyche Anderson, 
	and Jing Sun, \emph{Fellow, IEEE}

	\thanks{
		Qinan Zhou is with the Department of Mechanical Engineering, University of Michigan, Ann Arbor, MI, 48103, USA. (email: qinan@umich.edu).
        
		Gabrielle Vuylsteke and R. Dyche Anderson are with Research and Advanced Engineering, Ford Motor Company, Dearborn, MI 48124, USA. (email: gvuylste@ford.com, rander34@ford.com). 
		
		Jing Sun is with the Department of Naval Architecture and Marine Engineering, University of Michigan, Ann Arbor, MI 48103, USA. (email: jingsun@umich.edu). 
	}
}

\maketitle

\begin{abstract}
Incremental capacity analysis (ICA) and differential voltage analysis (DVA) are two effective approaches for battery degradation monitoring. One limiting factor for their real-world application is that they require constant-current (CC) charging profiles. This research removes this limitation and proposes an approach that extends ICA/DVA-based degradation monitoring from CC charging to dynamic charging profiles. A novel concept of virtual incremental capacity (VIC) and virtual differential voltage (VDV) is proposed. Then, two related convolutional neural networks (CNNs), called U-Net and Conv-Net, are proposed to construct  VIC/VDV curves and estimate the state of health (SOH) from dynamic charging profiles across any state-of-charge (SOC) range that satisfies some constraints. Finally, two CNNs called Mobile U-Net and Mobile-Net are proposed as replacements for the U-Net and Conv-Net, respectively, to reduce the computational footprint and memory requirements, while keeping similar performance. Using an extensive experimental dataset of battery modules, the proposed CNNs are demonstrated to provide accurate VIC/VDV curves and enable ICA/DVA-based battery degradation monitoring under various fast-charging protocols and different SOC ranges. 
\end{abstract}


\markboth{}%
{}

\definecolor{limegreen}{rgb}{0.2, 0.8, 0.2}
\definecolor{forestgreen}{rgb}{0.13, 0.55, 0.13}
\definecolor{greenhtml}{rgb}{0.0, 0.5, 0.0}

\section{INTRODUCTION}
\label{Intro}
\IEEEPARstart{E}{stimating} battery state of health (SOH) and monitoring associated degradation from onboard measurements are crucial for range estimation, performance, safety, maintenance, and warranty of electrified vehicles \cite{Noura}. SOH can be characterized by either capacity fading or resistance rising, both of which lead to a reduction in usable energy \cite{Berecibar,CalRegulation}. This paper focuses on capacity fading and defines SOH as $\text{SOH} = C/C_{\text{fresh}}$, where $C$ and $C_{\text{fresh}}$ are the capacities of a battery at its current and fresh state, respectively.

The battery degradation status can be assessed using intrusive and non-intrusive methods. Various intrusive methods, including photoelectron spectroscopy \cite{Philippe}, scanning electron microscopy \cite{Abellan}, X-ray diffraction \cite{Luo}, etc., provide the exact health and degradation status of batteries, but require expensive equipment and are more suitable for laboratory testing. Moreover, they may lead to changes in battery performance or even cause permanent damage \cite{Berecibar}. In contrast, non-intrusive methods use measurements such as current, voltage, and temperature to infer the battery degradation status. Popular non-intrusive methods include electrochemical impedance spectroscopy \cite{Li2025}, incremental capacity analysis (ICA) \cite{Dubarry, Xu}, differential voltage analysis (DVA) \cite{Xu2,Wang2025}, adaptive filtering \cite{PlettEKF,PlettSPKF, Nguyen}, and data-driven techniques \cite{Yang,Teixeira}. These methods do not require special instrumentation and can be implemented with existing onboard vehicle sensors.

This paper also focuses on ICA and DVA, two approaches that integrate physical insights with data analytics to interpret battery degradation from measured current and voltage profiles \cite{Weng1, Krupp, Dubarry}. Specifically, ICA and DVA yield incremental capacity (IC) and differential voltage (DV) curves, from which IC/DV features can be extracted \cite{Weng1}. Changes in these features reflect underlying degradation mechanisms within batteries \cite{Krupp, Dubarry}. The ability to extract such physical insights without complex models is a key strength of ICA/DVA compared to purely data-driven methods. Previous studies have demonstrated that IC/DV features can be used to accurately estimate cell-level SOH under different charging conditions \cite{Weng1,Weng2,Weng3,Zhou1,Stephens} and module-level SOH under cell-to-cell variations \cite{Zhou2}.

Although originally developed for open-circuit conditions \cite{Dubarry}, ICA and DVA have been adapted for charging under nonzero constant C-rate conditions. \cite{Tian01} reconstructs open-circuit profiles from constant-current (CC) charging and performs ICA/DVA on the reconstructed profiles. Another type of approach applies ICA/DVA directly to CC charging profiles, where the resulting IC/DV curves and features exhibit a C-rate dependency \cite{Weng1, Zhou1, Wong, ZhouR, Chen}. Thus, C-rate is treated as an additional feature for degradation monitoring \cite{Fly}. Nonetheless, constant current (CC) remains a key requirement. 

The primary limitation of ICA/DVA is the requirement of CC charging over a broad SOC range \cite{Zhang, Xu, John, Li20252}, which is often unmet in real-world charging scenarios with varying SOC windows and nonconstant charging currents, especially in fast charging and constant-power charging \cite{Tomaszewska}. As a result, ICA/DVA-based degradation monitoring becomes inapplicable. To relax the requirement for a broad SOC range, Tian et al. \cite{Tian02} reconstructed entire CC charging profiles using only 10 minutes of partial data. ICA/DVA has been extended from CC charging to pulsed constant-current charging \cite{Tang,Wind,Gong2025}. To the best of our knowledge, extending ICA/DVA to more dynamic real-world charging current profiles, such as fast charging profiles, has not been addressed in the literature.

This paper extends ICA/DVA to dynamic charging profiles\footnote{Hereafter, the term ``dynamic current profiles'' refers to current profiles that are not necessarily constant, and the term ``dynamic charging profiles'' denotes profiles that can be measured from charging with dynamic current profiles, such as profiles of voltage, current, transferred charge, etc.\label{shared_footnote}}. In contrast to prior studies \cite{Tian01, Tian02, Tang,Wind,Gong2025}, the proposed approach, for the first time, enables ICA/DVA-based SOH estimation and degradation monitoring without requiring constant-current conditions. To evaluate the performance of the approach, a large experimental dataset of battery modules with three cells connected in parallel is used. Specifically, the contributions of the paper are four-fold: 
\begin{itemize}
    \item First, notions of virtual IC/DV (VIC/VDV) curves are proposed to enable the extension of the conventional ICA/DVA-based SOH estimation approach from CC charging to dynamic charging. For cells or modules charged using dynamic current profiles\hyperref[shared_footnote]{\footnotemark[1]}, VIC/VDV curves are defined as the IC/DV curves that the same cells or modules, in the identical state of degradation, would exhibit under CC charging at a reference C-rate of interest.
    \item Second, a convolutional neural network (CNN), called U-Net, is developed to construct VIC/VDV curves from dynamic charging profiles with any SOC ranges satisfying some constraints. Applied to the aforementioned dataset, VIC/VDV curves are shown to approximate actual IC/DV curves very well. As a case study, 0.7\% root-mean-square error (RMSE) for SOH estimation is achieved using only two features extracted from the VIC/VDV curves.
    \item Third, a CNN called Conv-Net is proposed to estimate SOH directly, bypassing the construction of VIC/VDV curves. For the same dataset, Conv-Net achieves a root-mean-square error (RMSE) of 0.6\% in SOH estimation from dynamic charging profiles. 
    \item Fourth, computationally efficient CNNs, referred to as Mobile U-Net and Mobile-Net, are developed to reduce the computational and memory loads. For the same dataset, the Mobile U-Net and Mobile-Net are shown to reduce the total number of neural-network parameters and floating-point operations by roughly 67\%, while keeping similar performance for VIC/VDV curve construction and SOH estimation. 
\end{itemize} 
Note that the main contribution of this paper is not the proposal of new degradation-related features or feature extraction methods. Instead, the key contribution lies in extending the conventional ICA/DVA-based degradation monitoring framework, which is traditionally developed for CC charging, to dynamic charging conditions with time-varying current profiles.

To elucidate the proposed approach, the paper is organized as follows. Section \ref{ICA} defines VIC/VDV curves and features. Sections \ref{IC/DV_Construction} and \ref{Conv-Net} describe the proposed methods for VIC/VDV curve construction and direct SOH estimation, respectively, using dynamic charging profiles. Section \ref{DataAndPrep} discusses the battery module dataset used in this study. Section \ref{Result} assesses the performance of the proposed methods using experimental data. Section \ref{Conclusion} summarizes the paper.
\section{VIRTUAL INCREMENTAL CAPACITY AND DIFFERENTIAL VOLTAGE CURVES}
\label{ICA}
This section defines the proposed concepts that underpin this work. First, VIC/VDV curves are defined. Then, VIC/VDV features are briefly discussed.

IC/DV curves are originally defined in fully relaxed open-circuit conditions \cite{Dubarry}, which are rarely encountered in practice. To accommodate real-world applications, subsequent studies \cite{Weng1, Zhou1, Wong, ZhouR, Chen} have defined IC/DV curves as:
\begin{itemize}
    \item \textbf{IC Curve}: $\text{IC} = dQ_{\text{CC}}/dV_{\text{CC}}$ as a function of $V_{\text{CC}}$,
    \item \textbf{DV Curve}: $\text{DV} = dV_{\text{CC}}/dQ_{\text{CC}}$ as a function of $Q_{\text{CC}}$,
\end{itemize}
where $V_{\text{CC}}$ is voltage and $Q_{\text{CC}}$ is the charged capacity under CC charging with a reference C-rate of interest. The subscript ``CC'' emphasizes that these quantities are obtained from the constant-current conditions, and the shapes of IC/DV curves depend on the C-rate \cite{Fly}.

Consider battery cells or modules charged with dynamic current profiles. While $Q$-$V$ curves can still be measured and differentiated, the resulting derivative curves do not conform to the above definitions of IC/DV curves and therefore lack meaningful interpretation for ICA/DVA.

To overcome the reliance on CC charging and to extend the ICA/DVA-based degradation monitoring to dynamic charging conditions, this paper defines the VIC/VDV curves as the IC/DV curves that the same cells or modules, in the identical state of degradation, would exhibit under CC charging at the same reference C-rate, denoted as: 
\begin{itemize}
    \item \textbf{Virtual IC Curve}: $\widehat{\text{IC}}$ as a function of $\widehat{V}_{\text{CC}}$,
    \item \textbf{Virtual DV Curve}: $\widehat{\text{DV}}$ as a function of $\widehat{Q}_{\text{CC}}$,
\end{itemize}
where $\widehat{V}_{\text{CC}}$, $\widehat{Q}_{\text{CC}}$, $\widehat{\text{IC}}$, and  $\widehat{\text{DV}}$ are estimates that approximate the expected response under CC charging. The notation $\widehat{\cdot}$ emphasizes that they are inferred from charging profiles that are not necessarily constant-current.

Features extracted from VIC/VDV curves are called VIC/VDV features and defined analogously to IC/DV features commonly used in the literature \cite{Dubarry,Krupp,Bloom,Ansean,ZhouR}. The specific VIC/VDV features employed for SOH estimation in this work are detailed in Section \ref{Dataset}.
\section{VIRTUAL IC/DV CURVE CONSTRUCTION}
\label{IC/DV_Construction}
With key concepts defined, the next step is to develop a mapping from dynamic charging profiles to VIC/VDV curves so that ICA/DVA can be performed. Given the complexity of formulating such a mapping analytically, a CNN-based approach is adopted. Compared to other neural networks, CNN is selected for its strong ability to capture spatial information. Note that important IC/DV features, such as IC peak heights, locations, and areas, are all spatial information.

This section elaborates on the proposed method summarized in Algorithm \ref{alg:Method1}. The CNNs are implemented and evaluated using the Keras framework \cite{Keras}.

Note that the proposed CNNs are black-box models without embedding physical knowledge. However, once their construction accuracy is established by proper design and training, the virtual IC/DV curves can closely approximate the actual IC/DV curves and, therefore, serve as a reliable alternative when the latter cannot be obtained. As such, features visualized using proper tools, can be used to derive relevant physical interpretation, as shown in \cite{Erhan,raghakotkerasvis}.

\begin{algorithm}[h]
\textbf{Output}: $\left\{ \widehat{Q}_{\text{CC}} ( k ) \right\}$, $\left\{ \widehat{V}_{\text{CC}} ( k ) \right\}$, $\left\{ \widehat{\text{IC}} ( k ) \right\}$, $\left\{ \widehat{\text{DV}} ( k ) \right\}$ \

\textbf{Input}: $\left\{I\left( \Delta Q \right)\right\}$ and $\left\{V\left( \Delta Q \right)\right\}$ from dynamic charging, a downsampling increment $\Delta q$  \

\

\begin{algorithmic}[1]
\State $\left\{\tilde{I}\left( \Delta Q \right)\right\}$, $\left\{\tilde{V}\left( \Delta Q \right)\right\} \leftarrow$ Downsample 
\State $\left\{\tilde{I}\left( k \right)\right\}$, $\left\{\tilde{V}\left( k \right)\right\} \leftarrow$ Pad $\left\{\tilde{I}\left( \Delta Q \right)\right\}$ and $\left\{\tilde{V}\left( \Delta Q \right)\right\}$ 
\State $\left\{\tilde{I}^{\left( s \right)}\left( k \right)\right\}$, $\left\{\tilde{V}^{\left( s \right)}\left( k \right)\right\} \leftarrow$ Equation (\ref{eqn:ConvertStandardize}) 
\State $\left\{ \widehat{Q}^{\left( s \right)}_{\text{CC}} ( k ) \right\}$, $\left\{ \widehat{V}^{\left( s \right)}_{\text{CC}} ( k ) \right\}$, $\left\{ \widehat{\text{IC}}^{\left( s \right)} ( k ) \right\} \leftarrow $ CNN$\left(\left\{\tilde{I}^{\left( s \right)}\left( k \right)\right\}, \left\{\tilde{V}^{\left( s \right)}\left( k \right)\right\}\right)$ 
\State $\left\{ \widehat{Q}_{\text{CC}} ( k ) \right\}$, $\left\{ \widehat{V}_{\text{CC}} ( k ) \right\}$, $\left\{ \widehat{\text{IC}} ( k ) \right\} \leftarrow$ Equation (\ref{eqn:ConvertStandardize}) 
\State $\left\{ \widehat{\text{DV}} ( k ) \right\} \leftarrow 1/\left\{ \widehat{\text{IC}} ( k ) \right\}$ \ 
\end{algorithmic}

\
    
\tcc{$\Delta Q \in \left[0,\overline{\Delta Q}\right]$, $k\in\{1,2,...,N_{\text{nn}}\}$}  
 
\caption{To Construct VIC/VDV Curves}
\label{alg:Method1}
\end{algorithm}


\subsection{Input}
\label{Input}
Define $\text{SOC}_{\text{initial}}$, $\text{SOC}_{\text{final}}$, $\Delta\text{SOC} = \text{SOC}_{\text{final}} - \text{SOC}_{\text{initial}}$ as initial SOC, final SOC, and partial charging range, respectively. Consider a dynamic charging event that satisfy the following SOC window:
\begin{equation}
\label{eqn:SOCwindow}
\Delta\text{SOC} \geq \underline{\Delta\text{SOC}},
\end{equation} 
where the underline $\underline{\cdot}$ and overline $\overline{\cdot}$ represent the lower and upper bounds, respectively. As the SOC window narrows, charging profiles contain less information, leading to higher construction errors. Hence, for a given target accuracy, the smallest possible SOC window is selected by trial and error to maximize coverage across diverse SOC ranges. Moreover, since the degradation process evolves gradually, SOH estimation does not need to be performed for every charging cycle. Therefore, it is acceptable if an encountered SOC range occasionally falls beyond the chosen SOC window.

Let $\left\{I\left(t\right)\right\}$, $\left\{V\left(t\right)\right\}$, $\left\{\Delta Q\left(t\right)\right\}$, $t\in \left[0,T\right]$ be the profiles of current, voltage, and transferred charge over the interval $\left[0,T\right]$, respectively, where $t$ is time and $T$ is the total charging time. The transferred charge is defined as $\Delta Q \left( t \right) = \int_{0}^{t} I \left( \tau \right) d\tau \in \left[ 0, \overline{\Delta Q} \right] \label{eqn:CoulombCounting}$. $\left\{\Delta Q\left(t\right)\right\}$ always starts at 0 and ends at the total amount of transferred charge ($\overline{\Delta Q}$). Furthermore, both $\left\{I\left(t\right)\right\}$ and $\left\{V \left(t\right)\right\}$ can be re-written as functions of $\left\{\Delta Q\left(t\right)\right\}$, denoted as $\left\{I\left( \Delta Q \right)\right\}$, $\left\{V\left( \Delta Q \right)\right\}$, $\Delta Q \in \left[0,\overline{\Delta Q}\right]$, respectively. 

The inputs to the proposed method given in Algorithm \ref{alg:Method1} are $\left\{I\left( \Delta Q \right)\right\}$ and $\left\{V\left( \Delta Q \right)\right\}$. Hence, SOC is not used as an input.

\subsection{Downsampling}
\label{Downsampling}
$\left\{I\left( \Delta Q \right)\right\}$ and $\left\{V\left( \Delta Q \right)\right\}$ profiles measured onboard, even for a limited SOC range, typically have different sequence lengths that exceed the length required by the proposed CNNs. To standardize the inputs, the downsampling and the one-sided symmetric padding (to be explained in Section \ref{SymmetricPadding}) are applied, as illustrated in Algorithm \ref{alg:Method1}.

This paper proposes downsampling based on transferred charge ($\Delta Q$), rather than time or SOC, for the following reasons. First, time-based downsampling can misrepresent key regions of a charging profile. For instance, the segment from 80\% to 100\% SOC in fast charging has lower current and longer duration, leading to disproportionately higher sampling density in a small portion of the entire charging range \cite{Tomaszewska}. Second, while both SOC-based and $\Delta Q$-based downsampling can mitigate this issue, SOC must be estimated in practice \cite{PlettBook}, bringing additional uncertainty. Thus, downsampling based on $\Delta Q$ is the most suitable.

Specifically, $\left\{I\left( \Delta Q \right)\right\}$ and $\left\{V\left( \Delta Q \right)\right\}$ are downsampled for every incremental $\Delta q$ defined as: 
\begin{equation}
    \Delta q = \frac{ \Delta Q_{\text{max}} }{ N_{\text{nn}} }, \label{eqn:DownsamplingIncrement}
\end{equation}
where $N_{\text{nn}}$ is the required input sequence length of the proposed CNNs, and $\Delta Q_{\text{max}}$ is the maximum amount of transferred charge the proposed method needs to handle. One way to determine $\Delta Q_{\text{max}}$ is:
\begin{equation}
    \Delta Q_{\text{max}} = \overline{\Delta \text{SOC}} \cdot C_{\text{fresh}}, \label{eqn:DeltaQ_max}
\end{equation}
where $\overline{\Delta \text{SOC}}$ is the maximum charging SOC range to be handled, and $C_{\text{fresh}}$ is the capacity of a battery at the fresh state. Both $\Delta Q_{\text{max}}$ and $\Delta q$ are computed once during algorithm calibration and stored. Hence, no SOC information is needed for the algorithm implementation. 

\begin{remark}
     For each charging profile, the total amount of transferred charge $\overline{\Delta Q}$ is not necessarily divisible by $\Delta q$. Thus, after downsampling, each original profile often has a small residual with the amount of transferred charge less than $\Delta q$. Because $\Delta q$ is chosen to be small, this residual can be omitted for the rest of the proposed method.  
\end{remark}

\begin{figure*}[ht]
    \centering
    \includegraphics[width=0.7\textwidth,trim=10 60 4 15,clip]{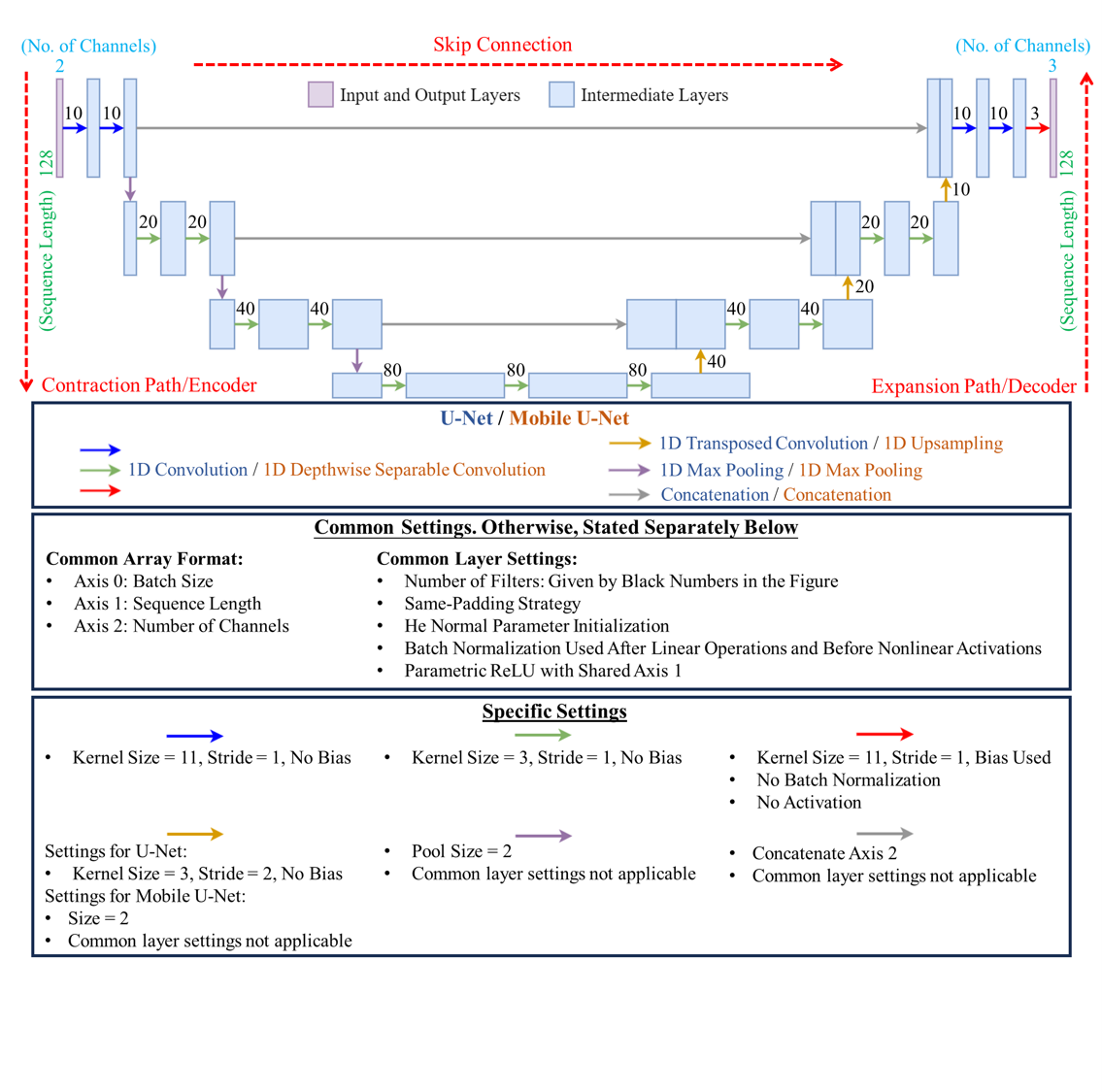}
    \caption{Designs of the Proposed U-Net (Section \ref{U_Net}) and Mobile U-Net (Section \ref{Mobile_U_Net}).}
    \label{fig:UNet}
\end{figure*} 

\subsection{One-Sided Symmetric Padding}
\label{SymmetricPadding}
Recall that the proposed CNNs require input profiles of a fixed sequence length ($N_{\text{nn}}$) \cite{Siddique}. Since original profiles have different $\overline{\Delta Q}$ and are downsampled with a fixed $\Delta q$, sequence lengths of downsampled profiles ($N_{\text{point}}$) are different for different profiles and $N_{\text{point}} \leq N_{\text{nn}}$. Thus, padding is applied to resolve this mismatch in input sequence length. 

Among various padding strategies \cite{Innamorati}, this paper modifies the strategy from \cite{Fan} because of its ability to reduce errors at the boundaries of the constructed profiles. Specifically, the one-sided symmetric padding scheme is used, where mirrored values are appended to the end of the target profile to reach the required length. For example, extending a profile array $\left[a, b, c\right]$ to a length of 8 yields $\left[a, b, c, c, b, a, a, b\right]$, where the original values remain at the beginning and mirrored values are appended at the end. In general, a profile with $N_{\text{point}}$ points is padded with $(N_{\text{nn}} - N_{\text{point}})$ mirrored points at the end to produce a final length of $N_{\text{nn}}$.

After downsampling and padding, the resulting current and voltage profiles, denoted as $\left\{\tilde{I}\left( k \right)\right\}$, $\left\{\tilde{V}\left( k \right)\right\}$, $k\in\{1,2,...,N_{\text{nn}}\}$, respectively, are sent as two separate channels to the proposed CNNs, where $k$ is an index. $\Delta Q$ profile is no longer needed, as it is implicitly embedded in the downsampled current and voltage profiles.

\subsection{Output}
\label{Output}
The outputs of the proposed CNNs contain three separate channels: $\left\{ \widehat{Q}_{\text{CC}} ( k ) \right\}$, $\left\{ \widehat{V}_{\text{CC}} ( k ) \right\}$, $\left\{ \widehat{\text{IC}} ( k ) \right\}$, $k\in \left\{1,2,...,N_{\text{nn}}\right\}$. $\left\{ \widehat{\text{DV}} ( k ) \right\}$ is computed as $\widehat{\text{DV}}( k ) = 1/\widehat{\text{IC}}( k )$. The output SOC range is fixed and the same for all predictions, regardless of the input SOC range. While this fixed range does not need to be full-range, it must be broad enough to include all the IC/DV features of interest. This paper denotes this fixed SOC range as $\widehat{\text{SOC}} \in \left[ \widehat{\text{SOC}}_{\text{initial}}, \widehat{\text{SOC}}_{\text{final}} \right]$. Within this range, datapoints for output profiles are evenly distributed. Due to the symmetry of the U-Net architecture (to be discussed in Section \ref{U_Net}), the output sequence length is also $N_{\text{nn}}$, equal to the input sequence length. 

\subsection{U-Net for VIC/VDV Curve Construction}
\label{U_Net}
This subsection elaborates on the proposed CNN, called U-Net. The details of the proposed U-Net are shown in Fig. \ref{fig:UNet}. U-Net is selected over other CNNs for two reasons. First, its encoder-decoder structure provides a clear separation between feature extraction and profile construction, enhancing explainability \cite{Siddique}. Second, its skip connections preserve spatial information at both local and global levels, allowing the CNN to integrate multi-scale features for more accurate profile construction \cite{Azad}.

The U-Net shown in Fig. \ref{fig:UNet} consists of three main components: the contraction path, skip connections, and expansion path \cite{Ronneberger}. The contraction path (or encoder), composed of convolution and max pooling, extracts features at different levels \cite{Siddique}. As its depth increases, extracted features become more global and less local \cite{Siddique}. Skip connections (or concatenations) send the extracted features to the expansion path, allowing CNN to leverage both local and global information \cite{Azad}. The expansion path (or decoder), consisting of convolution and transposed convolution, constructs profiles using received features \cite{Siddique}. Transposed convolution optimally increases the resolution of constructed profiles, while convolution combines the constructed profiles with received features \cite{Ronneberger}.

U-Net offers significant design flexibility, including the number of levels and layers, as well as layer-specific configurations. Fig. \ref{fig:UNet} shows the specific design used by this paper. While some design choices are directly adopted from \cite{Ronneberger}, others are customized to fit the needs of this study and are worthy of discussion. 
\begin{itemize}
    \item \textbf{Performance Enhancement}: First, larger kernel sizes are used in early and late layers to suppress input noise in the contraction path and generate smoother output profiles in the expansion path. Second, parametric rectified linear unit (PReLU) \cite{He} is used for activation. It learns optimal leakage hyperparameters to solve the dying ReLU problem, thereby leading to better performance without an increase in computational complexity \cite{He}. As in \cite{He}, the PReLU is applied with a shared leakage hyperparameter per channel and distinct hyperparameters across channels. 
    \item \textbf{Improved Training Efficiency}: He Normal parameter initialization \cite{He} and batch normalization \cite{Ioffe} are used to mitigate vanishing gradient problems \cite{CNN_Book}. Following \cite{Ioffe}, batch normalization is applied after linear operations and before nonlinear activations, effectively removing the effects of biases in linear operations. The default hyperparameters in Keras \cite{Keras} are adopted for batch normalization.
\end{itemize}

\subsection{Standardization and Training}
\label{U-Net-Training}
To ensure unbiased training and faster convergence, all the input profiles ($\left\{\tilde{I}\left( k \right)\right\}$, $\left\{\tilde{V}\left( k \right)\right\}$) and output profiles ($\left\{ \widehat{Q}_{\text{CC}} ( k ) \right\}$, $\left\{ \widehat{V}_{\text{CC}} ( k ) \right\}$, $\left\{ \widehat{\text{IC}} ( k ) \right\}$) are standardized using: 
\begin{equation}
\left\{ z^{\left( s \right)} ( k ) \right\} = \frac{{\left\{ z ( k ) \right\} } - \mu_z}{\sigma_z}, \label{eqn:ConvertStandardize}
\end{equation}
where $z$ can be any of the variables listed above, and $\mu_z$ and $\sigma_z$ are the mean and standard deviation computed over all the points in all the training samples:
\begin{eqnarray}
    \mu_z &=& \text{mean} \left( \left\{\left[\left\{ z ( k ) \right\} \right]_j \right\}_{j=1}^{N} \right), \\ 
     \sigma_z &=& \text{std} \left( \left\{\left[\left\{ z ( k ) \right\} \right]_j \right\}_{j=1}^{N} \right), 
\end{eqnarray}
where $\left[\left\{ z ( k ) \right\} \right]_j$ denotes the profile $\left\{ z ( k ) \right\}$ from the $j$-th training sample, $k\in\{1,2,...,N_{\text{nn}}\}$, and $N$ is the total number of training samples.

\begin{figure*}[ht]
    \centering
    \includegraphics[width=0.7\textwidth,trim=9 80 5 8,clip]{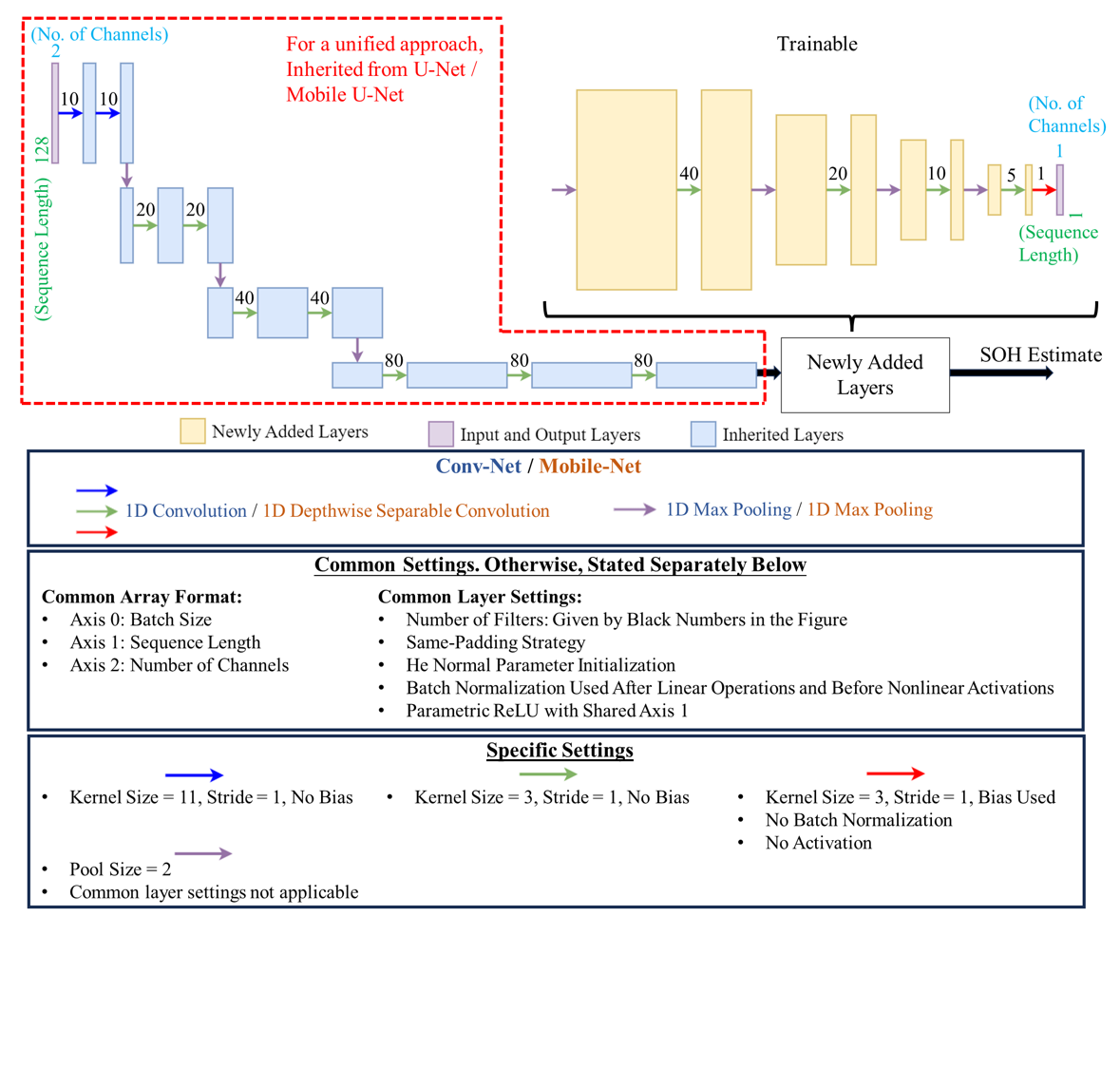}
    \caption{Designs of the Proposed Conv-Net and Mobile-Net}
    \label{fig:ConvNet}
\end{figure*}

The mean squared error (MSE) loss is used in this paper. The ADAM solver \cite{Kingma} with default settings in Keras \cite{Keras} is used to solve the optimization problem, allowing adaptive learning rate, faster convergence, and bias corrections. As a regularization technique, early stopping is employed to halt the training process when the validation MSE loss does not decrease for consecutive $N_{\text{th}}$ epochs \cite{CNN_Book}. 

\subsection{Mobile U-Net for Computational Efficiency}
\label{Mobile_U_Net}
The computational and memory demands of the proposed U-Net can be further reduced without compromising performance. Inspired by \cite{Howard}, this paper derives a more computationally efficient CNN called Mobile U-Net to replace U-Net for onboard implementations.

As shown in Fig. \ref{fig:UNet}, the proposed Mobile U-Net retains the same architecture as the U-Net, but incorporates two key modifications. First, depthwise separable convolution is used in place of convolution. Originally proposed in \cite{Howard} and widely adopted in CNNs for mobile and embedded applications \cite{Howard, Sandler}, depthwise separable convolution consists of a depthwise convolution for spatial filtering followed by a pointwise convolution for channel-wise combination. This structure significantly reduces both computation and memory cost compared to conventional convolutions \cite{Howard}. Second, upsampling with simple repetition (as implemented in Keras \cite{Keras}) replaces transposed convolution. Such upsampling has no parameters to be learned. Compared to transposed convolution, upsampling loses the ability to learn optimal upsampling strategies, but offers substantial savings in computation and memory. A detailed comparison of the computation and memory for U-Net and Mobile U-Net will be given in Section \ref{OnboardImplementation}. 

Otherwise, as shown in Fig. \ref{fig:UNet}, Mobile U-Net employs the same layer design strategies and training procedure described in Sections \ref{U_Net} and \ref{U-Net-Training}, respectively.
\section{DIRECT SOH ESTIMATION}
\label{Conv-Net} 
With the trained U-Net and Mobile U-Net, general-purpose VIC/VDV curves can be constructed for various degradation-related tasks, such as assessing remaining useful life \cite{Pang} and detecting internal short-circuit faults \cite{Zhao}. However, if SOH estimation is the sole objective, these CNNs can be simplified to provide SOH estimates directly from dynamic charging profiles with any SOC ranges satisfying Constraint (\ref{eqn:SOCwindow}), while achieving higher estimation accuracy than using constructed VIC/VDV curves. The simplified CNNs are called Conv-Net and Mobile-Net.

While many CNNs are proposed in the literature for SOH estimation \cite{Yang, Gu, Chemali}, most of them use fully connected layers, which introduce large parameter counts and substantial computational overhead. In contrast, the proposed Conv-Net uses only convolutions, resulting in significantly fewer parameters and improved computational efficiency. By adopting depthwise separable convolutions, Mobile-Net is proposed to further simplify Conv-Net, while maintaining estimation performance. A detailed comparison of their computational footprints will be given in Section \ref{OnboardImplementation}.

The proposed method uses the same input, downsampling, and one-sided symmetric padding as what is discussed in Sections \ref{Input}, \ref{Downsampling}, and \ref{SymmetricPadding}. The outputs are SOH estimates ranging from 0 to 1. The overall procedure of the proposed method is summarized in Algorithm \ref{alg:Method2}.  

\begin{algorithm}[h]
\textbf{Output}: SOH estimate $\widehat{\text{SOH}} \in [0,1]$ \

\textbf{Input}: $\left\{I\left( \Delta Q \right)\right\}$ and $\left\{V\left( \Delta Q \right)\right\}$ from dynamic charging, a pre-stored increment $\Delta q$  \

\

\begin{algorithmic}[1]
\State $\left\{\tilde{I}\left( \Delta Q \right)\right\}$, $\left\{\tilde{V}\left( \Delta Q \right)\right\} \leftarrow$ Downsample 
\State $\left\{\tilde{I}\left( k \right)\right\}$, $\left\{\tilde{V}\left( k \right)\right\} \leftarrow$ Pad $\left\{\tilde{I}\left( \Delta Q \right)\right\}$ and $\left\{\tilde{V}\left( \Delta Q \right)\right\}$ 
\State $\left\{\tilde{I}^{\left( s \right)}\left( k \right)\right\}$, $\left\{\tilde{V}^{\left( s \right)}\left( k \right)\right\} \leftarrow$ Equation (\ref{eqn:ConvertStandardize}) \
\State $\widehat{\text{SOH}} \leftarrow $  CNN$\left(\left\{\tilde{I}^{\left( s \right)}\left( k \right)\right\}, \left\{\tilde{V}^{\left( s \right)}\left( k \right)\right\}\right)$ \
\end{algorithmic} 

\

   \tcc{$\Delta Q \in \left[0,\overline{\Delta Q}\right]$, $k\in\{1,2,...,N_{\text{nn}}\}$ } 
\caption{To Construct SOH Estimates}
\label{alg:Method2}
\end{algorithm}

Specific designs for Conv-Net and Mobile-Net are shown in Fig. \ref{fig:ConvNet}. It should be noted that these models can be trained from scratch, independently of the U-Net and Mobile U-Net. However, to maintain a unified framework across VIC/VDV construction and SOH estimation, transfer learning is employed. Specifically, the entire contraction path of the trained U-Net and Mobile U-Net is transferred to the Conv-Net and Mobile-Net, respectively, for feature extraction. Since the contraction path extracts degradation-related features for VIC/VDV construction, it is well-suited for direct SOH prediction. During training, the parameters of the contraction path are frozen, and additional layers of convolution or depthwise separable convolution are appended for regression.

Similar to Section \ref{U-Net-Training}, the MSE loss, ADAM solver \cite{Kingma}, and early stopping are used for training. As SOH values lie in the range [0, 1], output standardization is unnecessary. However, inputs are still standardized using Equation (\ref{eqn:ConvertStandardize}) to ensure stable and fast convergence.

\begin{remark}
    (On Direct Construction of VIC/VDV Features) The proposed Conv-Net and Mobile-Net architectures can also be adapted to directly predict key VIC/VDV features, instead of VIC/VDV curves. In this case, the final layer of the CNNs shown in Fig. \ref{fig:ConvNet} can be configured with multiple output channels corresponding to the number of target VIC/VDV features, rather than a single channel used for SOH estimation. The training procedure follows the same settings described in Section \ref{U-Net-Training}. Both inputs and outputs are standardized according to Equation (\ref{eqn:ConvertStandardize}). 
\end{remark}
\begin{remark}
    (On Construction of VIC/VDV Curves vs Features) This work focuses on constructing complete VIC/VDV curves rather than specific VIC/VDV features, as the former offers greater generality and versatility. VIC/VDV curves can be utilized for a broad range of applications, including SOH estimation, remaining useful life prediction \cite{Pang}, and internal short-circuit fault detection \cite{Zhao}, which allows a single trained model to serve multiple diagnostic and prognostic purposes. In contrast, if models are designed to output specific VIC/VDV features, different models are required for different tasks because different features are informative for different objectives.
\end{remark}

\section{BATTERY DATASETS USED FOR EVALUATION}
\label{DataAndPrep}
This section describes the battery dataset used in this study for performance evaluation. Although the original dataset only contains a long charging range, random truncation of the charging profiles is applied to generate randomly varying SOC ranges, thereby better reflecting real-world partial charging scenarios. Section \ref{Dataset} introduces the original dataset, while Section \ref{Prep} details the random truncation process used to simulate diverse partial charging conditions.

\subsection{Original Battery Dataset}
\label{Dataset}
A proprietary dataset, comprising 96 lithium nickel-manganese-cobalt oxide 622 (NMC622) modules from a real electrified vehicle battery pack, is used. Each module consists of three NMC622 cells connected in parallel. Table \ref{table:Datasets} summarizes the key attributes of the dataset. Data were collected by real onboard sensors on a commercially available electrified vehicle at a 1 Hz sampling rate, without imposing any special-purpose sensing or sampling requirement beyond what is available on the vehicle. Since California regulations require electrified vehicle batteries to maintain at least 80\% of their original capacity \cite{CalRegulation}, this private dataset contains no samples with SOH values below 80\%. Consequently, the proposed methods are validated under SOH $\geq$ 80\%.

\begin{figure}[H]
    \centering
    \includegraphics[width=0.49\textwidth,trim=40 1 45 15,clip]{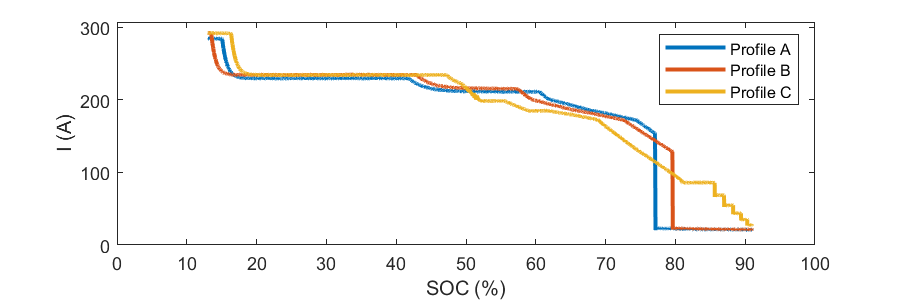}
\caption{Fast-Charging Current Profiles in the Dataset}
\label{fig:NonCC_Current_Profiles}
\end{figure}

These modules are cycled alternately using low C-rate and fast charging profiles. Rest periods are inserted before and after each cycle to allow full voltage relaxation, enabling accurate determination of both initial and final SOC, as well as module-level SOH. Within each module, three parallelly connected cells have self-balancing effects and do not involve any balancing circuitry. The low C-rate charging profile includes two CC charging stages with 86A (0.4C) and 64.5A (0.3C) module-level currents, and one subsequent constant-voltage stage. Module-level IC/DV curves are computed from the first CC stage, as this stage contains all important IC/DV features for this dataset. The analytical derivative method proposed by \cite{Weng1,Weng2,Weng3,Weng4} is used here to incorporate some robustness against high-frequency measurement noise. Additionally, this dataset contains three types of fast-charging protocols, as shown in Fig. \ref{fig:NonCC_Current_Profiles}. They are used to validate the feasibility of the proposed methods in extending ICA/DVA to dynamic charging profiles, which represents the primary objective of this study. However, due to the limitations of the current dataset, the performance of the proposed methods under other charging protocols with substantially different dynamic characteristics cannot yet be comprehensively evaluated. 

\begin{table}[H]
\centering
\caption{Key Attributes of the Battery Dataset}
\label{table:Datasets}
\begin{tabular}{|l|l|} 
\hline
\textbf{Attributes} & \textbf{Descriptions} \\ \hline\hline
Chemistry & NMC622 \\ \hline
Module Nominal Capacity & 208Ah \\ \hline
Module Configuration & 3 Cells in Parallel \\ \hline
No. of Modules & 96 \\ \hline
Module-Level SOH Range & 100\% - 86\% \\ \hline
No. of Fast Charging Protocols & 3 \\ \hline
No. of Input-Output Pairs & 40512 \\ \hline
No. of Input-Output Pairs Derived by Truncating & 405120 \\ \hline
\end{tabular}
\end{table}

\begin{figure}[H]
    \centering
    \includegraphics[width=0.49\textwidth,trim=25 90 65 65,clip]{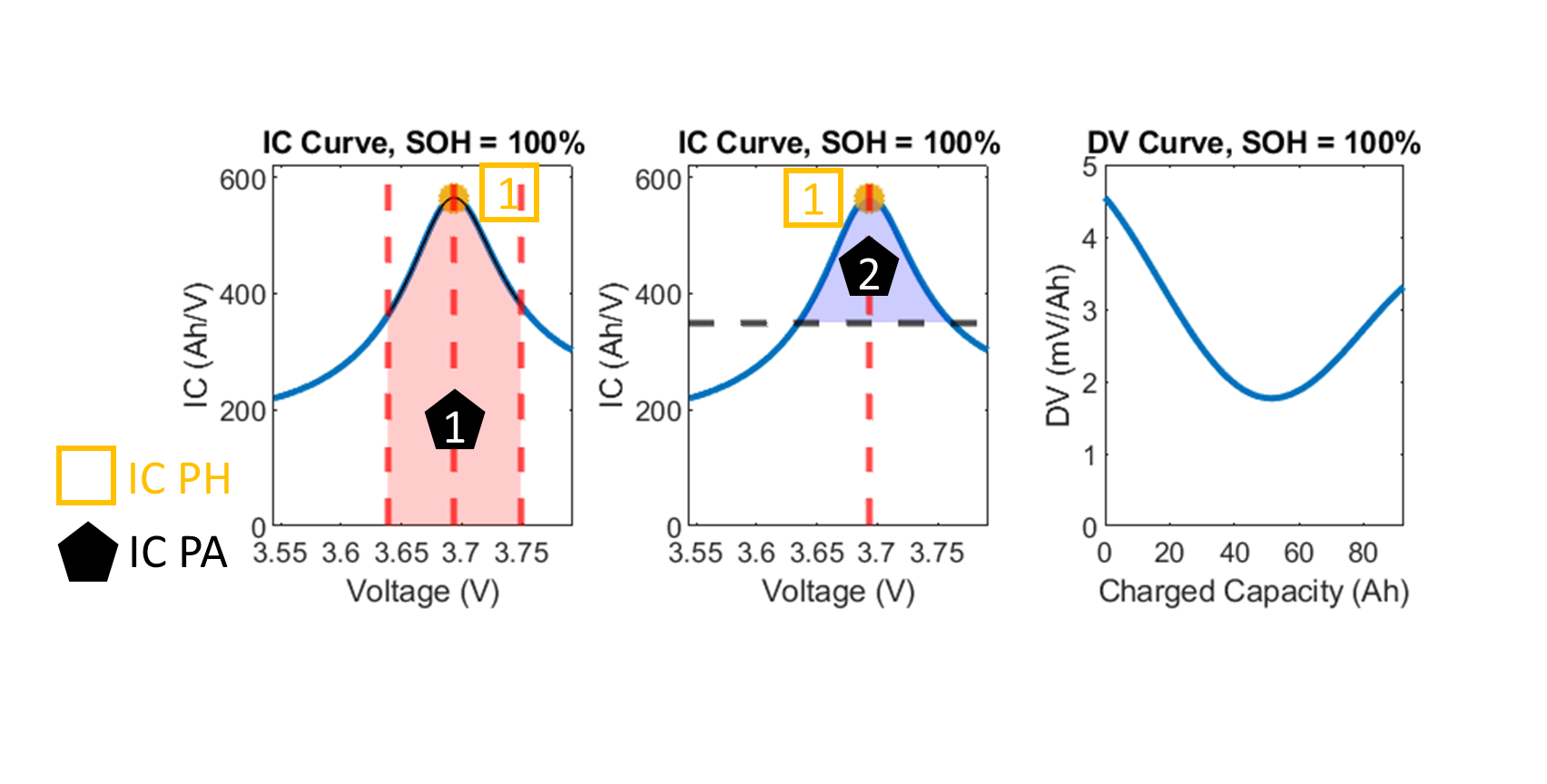}
\caption{Example IC/DV Curves and Features in the Dataset}
\label{fig:NMC622_Example_IC}
\end{figure}

Fig. \ref{fig:NMC622_Example_IC} shows one example of module-level IC/DV curves. This paper uses features such as IC peak height (IC PH) and IC partial areas (IC PA). As shown in Fig. \ref{fig:NMC622_Example_IC}, the IC PH is defined as the $y$-coordinate of the IC peak, and the IC PA is defined as the area either above a user-defined horizontal cutoff line (IC PA 2) or within a user-defined symmetric voltage window around a target IC peak (IC PA 1) \cite{ZhouR}. 

\begin{figure*}[ht]
    \centering
    \begin{subfigure}[h]{0.8\textwidth}
        \centering
        \includegraphics[width=\textwidth,trim=4 5 5 1,clip]{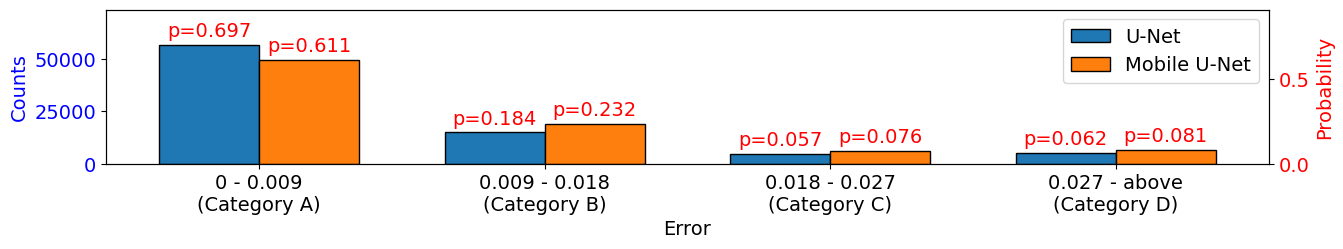}
        \caption{Error Distribution of the Testing Set}
        \label{fig:ModuleErrorDistribution}
    \end{subfigure} \hfill

    \begin{subfigure}[h]{0.19\textwidth}
        \centering
        \includegraphics[width=\textwidth,trim=1 5 1 1,clip]{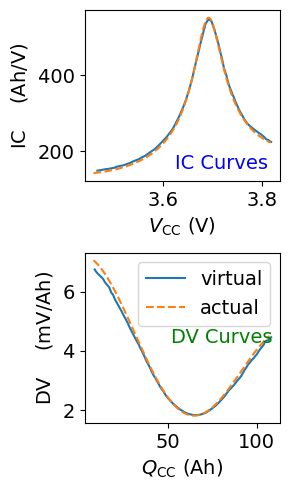}
        \caption{Category A}
        \label{fig:Module_CategoryA}
    \end{subfigure}
        \begin{subfigure}[h]{0.19\textwidth}
        \centering
        \includegraphics[width=\textwidth,trim=1 5 1 1,clip]{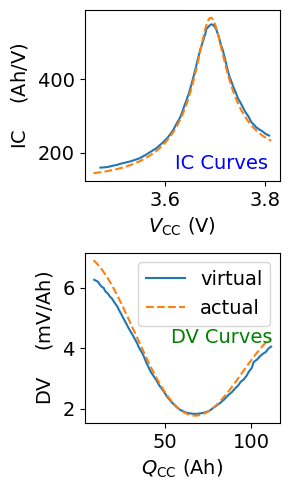}
        \caption{Category B}
        \label{fig:Module_CategoryB}
    \end{subfigure}
        \begin{subfigure}[h]{0.19\textwidth}
        \centering
        \includegraphics[width=\textwidth,trim=1 5 1 1,clip]{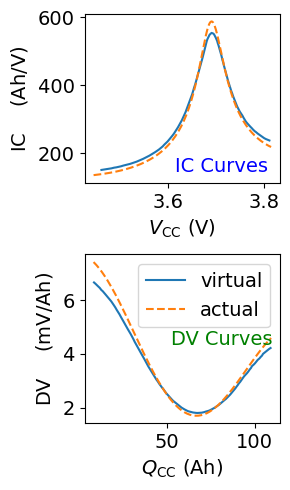}
        \caption{Category C}
        \label{fig:Module_CategoryC}
    \end{subfigure}
        \begin{subfigure}[h]{0.19\textwidth}
        \centering
        \includegraphics[width=\textwidth,trim=1 5 1 1,clip]{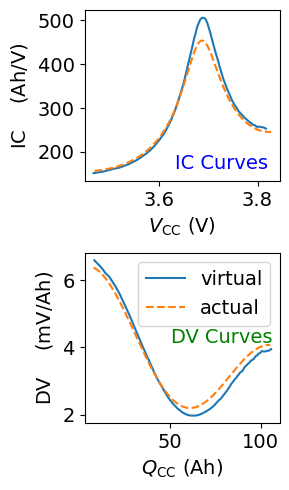}
        \caption{Category D}
        \label{fig:Module_CategoryD}
    \end{subfigure}
\caption{Performance of VIC/VDV Curve Construction under Random Input SOC Ranges}
\label{fig:Module_IC_Curve_Construction}
\end{figure*}

For training the proposed U-Net and Mobile U-Net, the module-level IC/DV curves from one low C-rate charging cycle are used as the ground truth for the outputs, while the module-level fast charging profiles from the next cycle are used as inputs.

\subsection{Dataset Used for Developing Proposed CNNs}
\label{Prep}
The original dataset contains fast-charging profiles with a SOC range spanning from 13\% to 91\%. To better reflect real-world conditions where the charging SOC range varies, these profiles are randomly truncated to generate a more diverse dataset with different SOC ranges satisfying Constraint (\ref{eqn:SOCwindow}). 

To train CNNs that generalize across arbitrary SOC ranges within the specified window, truncation must be performed in a randomized manner. Assuming equal importance across different SOC ranges, the sampling distribution should be a uniform distribution satisfying Constraint (\ref{eqn:SOCwindow}). The Dirichlet-rescale algorithm proposed by \cite{Griffin} is used to generate random SOC ranges from such a constrained uniform distribution. To further enrich the data, each input profile from the original dataset is truncated $N_{\text{truncate}}$ times to produce multiple profiles with distinct SOC ranges. The value of $N_{\text{truncate}}$ is chosen empirically based on the desired dataset size and computational budget. The number of input-output pairs after random truncations is given in Table \ref{table:Datasets}. 
\section{PERFORMANCE EVALUATION}
\label{Result}
This section evaluates the performance of the proposed CNNs using the dataset prepared in Section \ref{Prep}. The following specific case study is used as an example to illustrate the performance: $ \Delta\text{SOC} \geq 20\%$, $\overline{\Delta \text{SOC}} = 91\%-13\% = 78\%$, $N_{\text{nn}} = 128$, $\widehat{\text{SOC}}_{\text{initial}} = 5\%$, $\widehat{\text{SOC}}_{\text{final}} = 56\%$, $N_{\text{truncate}}=10$, $N_{\text{th}}=30$, and a mini-batch size of 64 for the ADAM solver. As explained in Section \ref{Input}, a trade-off exists between the SOC window and construction accuracy. Thus, the proposed methods can be further extended to $ \Delta\text{SOC} < 20\%$, but with deteriorated construction accuracy.

The dataset described in Section \ref{Prep} is randomly split into 60\% training, 20\% validation, and 20\% testing sets for development and evaluation. The comparisons among the proposed U-Net, Mobile U-Net, Conv-Net, and Mobile-Net are elucidated.

\subsection{Accuracy of Constructed VIC/VDV Curves}
\label{CurvesResult}
As noted in Sections \ref{Intro} and \ref{ICA}, derivatives directly computed from fast-charging profiles do not satisfy the definitions of IC/DV curves and, thus, are unsuitable for ICA/DVA. Hence, VIC/VDV curves are constructed as an alternative. This subsection evaluates the accuracy of module-level VIC/VDV curve construction using module-level fast charging profiles through the proposed U-Net and Mobile U-Net.

Recall that outputs of the proposed CNNs are $\left\{ \widehat{Q}^{\left(s\right)}_{\text{CC}} ( k ) \right\}$, $\left\{ \widehat{V}^{\left(s\right)}_{\text{CC}} ( k ) \right\}$, $\left\{ \widehat{\text{IC}}^{\left(s\right)} ( k ) \right\}$, which approximate $\left\{ Q^{\left(s\right)}_{\text{CC}} ( k ) \right\}$, $\left\{ V^{\left(s\right)}_{\text{CC}} ( k ) \right\}$, $\left\{ \text{IC}^{\left(s\right)} ( k ) \right\}$. The construction error can be defined as:
\begin{equation}
    \text{Error} = \frac{1}{N_{\text{nn}}} \sum_{k=1}^{N_{\text{nn}}} \left\|  \begin{bmatrix} \text{IC}^{\left(s\right)} ( k ) \\  V^{\left(s\right)}_{\text{CC}} ( k ) \\ Q^{\left(s\right)}_{\text{CC}} ( k ) \end{bmatrix} - \begin{bmatrix} \widehat{\text{IC}}^{\left(s\right)} ( k ) \\  \widehat{V}^{\left(s\right)}_{\text{CC}} ( k ) \\ \widehat{Q}^{\left(s\right)}_{\text{CC}} ( k ) \end{bmatrix} \right\|_2^2, \label{eqn:MSE}
\end{equation}
where $\| \cdot \|_2$ is the Euclidean norm. 

Fig. \ref{fig:ModuleErrorDistribution} shows the error distribution of the testing set under random input SOC ranges, divided into four categories. For each category, a pair of virtual and actual IC/DV curves from Mobile U-Net with an average error is shown in the rest of Fig. \ref{fig:Module_IC_Curve_Construction}. Several observations can be drawn. First, most VIC/VDV curves closely match their corresponding actual IC/DV curves, indicating that U-Net and Mobile U-Net can effectively construct accurate VIC/VDV curves for ICA/DVA-based degradation monitoring. Second, Mobile U-Net achieves similar performance to U-Net in VIC/VDV curve construction, but offers significantly lower computational requirements (to be discussed in Section \ref{OnboardImplementation}).

\begin{figure}[H]
    \centering
    \begin{subfigure}[h]{0.4\textwidth}
        \centering
        \includegraphics[width=\textwidth,trim=5 10 5 5,clip]{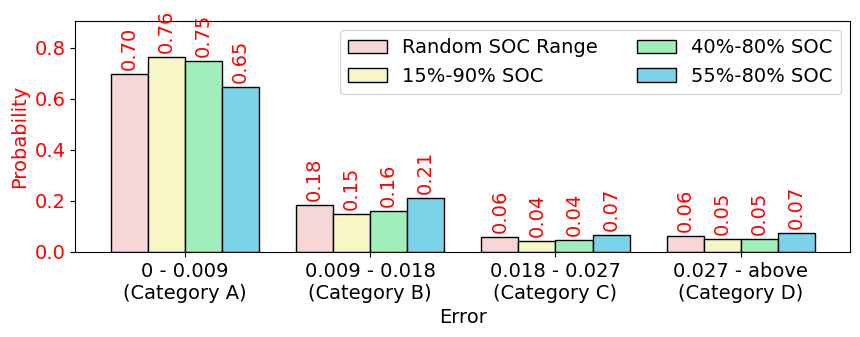}
        \caption{U-Net}
        \label{fig:Module_U_Net_Error_Distribution_CaseStudy}
    \end{subfigure} \hfill

    \begin{subfigure}[h]{0.4\textwidth}
        \centering
        \includegraphics[width=\textwidth,trim=5 10 5 5,clip]{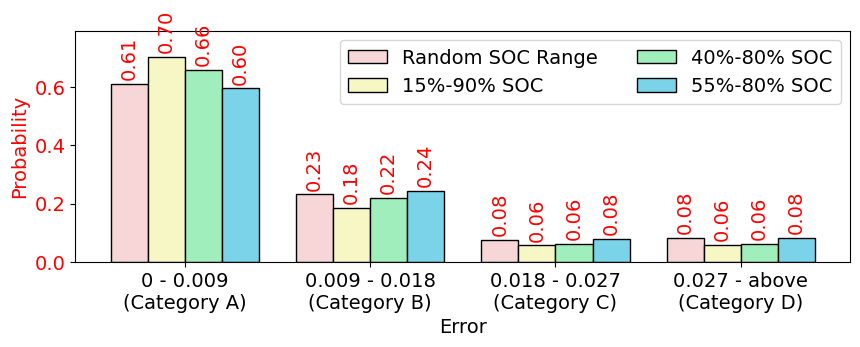}
        \caption{Mobile U-Net}
        \label{fig:Module_Mobile_U_Net_Error_Distribution_CaseStudy}
    \end{subfigure} 
\caption{Testing Error Distributions of VIC/VDV Curve Construction under Different Input SOC Ranges}
\label{fig:Module_Error_Distribution_Specific_Ranges}
\end{figure}

To evaluate the generalizability of the U-Net and Mobile U-Net trained from random input SOC ranges to specific input SOC ranges, the following three fast-charging cases are investigated: wide-range (15\%-90\% SOC), medium-range (40\%-80\% SOC), and narrow-range (55\%-80\% SOC). These test cases are generated by directly truncating profiles from the original dataset. The U-Net and Mobile U-Net are not re-trained for any specific ranges. Fig. \ref{fig:Module_Error_Distribution_Specific_Ranges} compares the error distributions of these cases against the random-range baseline. Results show that the performance of the VIC/VDV construction remains consistent across different specific ranges, with only slight deterioration when the SOC range is narrower due to reduced information content.

In summary, the proposed U-Net and Mobile U-Net enable ICA/DVA-based degradation monitoring from dynamic charging profiles, overcoming the constant-current limitations of traditional methods. Models trained on random SOC ranges generalize well to any fast-charging profiles satisfying the range constraints in Section \ref{Prep}.

\subsection{Applications of VIC/VDV Curves and Features}
\label{FeaturesResult}
To demonstrate the utility of the constructed module-level VIC/VDV curves for battery degradation monitoring, this paper uses module-level SOH estimation as a case study. 

For module-level SOH estimation, by using feature selection algorithms, an optimal set of IC/DV features can be found to combat against cell-to-cell variations \cite{Zhou2}. For the given dataset, IC partial areas (IC PA 1 and IC PA 2 defined in Fig. \ref{fig:NMC622_Example_IC}) are ranked the top two \cite{Zhou2} and are selected for this study. Then, using the relevance vector regression from \cite{Zhou2}, models are learned to map these features to SOH.

\begin{table}[H]
\centering
\caption{SOH Estimation Performance under Fast Charging Based on Testing Set}
\label{table:Estimation_Result_Table}
\begin{tabular}{|c|c|c|c|} 
\hline
\begin{tabular}[c]{@{}c@{}}\textbf{Neural}  \\ \textbf{Network} \end{tabular} & \textbf{Approach} & \textbf{RMSE} & \begin{tabular}[c]{@{}c@{}} \textbf{99.7th-Percentile} \\ \textbf{Absolute Error} \end{tabular} \\ 
\hline\hline
U-Net & \begin{tabular}[c]{@{}c@{}} Through VIC/VDV \\ Features \end{tabular} & \begin{tabular}[c]{@{}c@{}} 0.73\% \\ SOH \end{tabular} & 2.54\% SOH \\ 
\hline
Mobile U-Net & \begin{tabular}[c]{@{}c@{}} Through VIC/VDV \\ Features \end{tabular} & \begin{tabular}[c]{@{}c@{}} 0.79\% \\ SOH \end{tabular} & 2.86\% SOH \\ 
\hline
Conv-Net & \begin{tabular}[c]{@{}c@{}} Direct \\ Estimation \end{tabular} & \begin{tabular}[c]{@{}c@{}} 0.64\% \\ SOH \end{tabular} & 2.28\% SOH \\ 
\hline
Mobile-Net & \begin{tabular}[c]{@{}c@{}} Direct \\ Estimation \end{tabular} & \begin{tabular}[c]{@{}c@{}} 0.68\% \\ SOH \end{tabular} & 2.44\% SOH \\ 
\hline
\end{tabular}
\end{table}

The first two rows of Table \ref{table:Estimation_Result_Table} summarize the SOH estimation performance under fast charging with random input SOC ranges using VIC/VDV features. The extracted VIC/VDV features achieve a good performance well below the 5\% SOH error threshold required by California regulations \cite{CalRegulation}, demonstrating the effectiveness of the constructed VIC/VDV curves for degradation monitoring under dynamic charging. As to be further discussed in Section \ref{SOHResult}, VIC/VDV are more general-purpose (because different features can be extracted from VIC/VDV curves for different tasks), albeit with slightly reduced estimation accuracy compared to direct SOH estimation approaches.

\subsection{Accuracy of Direct SOH Estimates}
\label{SOHResult}
When SOH estimation is the sole objective, the proposed U-Net and Mobile U-Net can be simplified into Conv-Net and Mobile-Net, respectively. This subsection evaluates and compares the performance of Conv-Net and Mobile-Net in directly estimating SOH from dynamic charging profiles.

The last two rows of Table \ref{table:Estimation_Result_Table} summarize the SOH estimation performance using Conv-Net and Mobile-Net. The direct SOH estimation using Conv-Net and Mobile-Net outperforms the VIC/VDV feature-based SOH estimation using U-Net and Mobile U-Net. The performance also significantly surpasses California's regulatory requirements in Section \ref{FeaturesResult}. Hence, Conv-Net and Mobile-Net have the advantage of better SOH estimation performance, while U-Net and Mobile U-Net offer the complementary benefit of generating general-purpose VIC/VDV curvs whose features can be used for other tasks besides SOH estimation, such as the remaining useful life prediction \cite{Pang} and internal short-circuit fault detection \cite{Zhao}. 

\begin{figure}[H]
    \centering

    \includegraphics[width=0.45\textwidth,trim=5 10 5 5,clip]{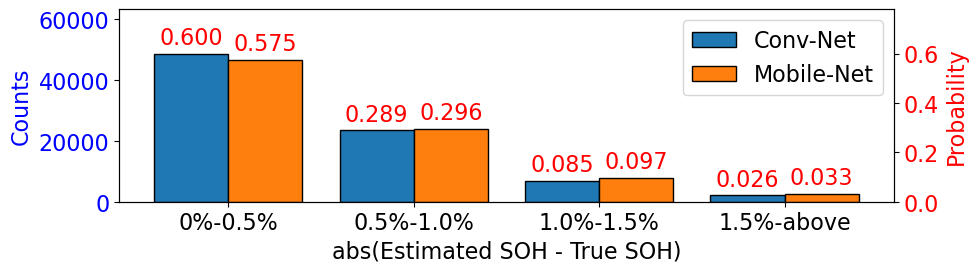}

\caption{Testing Absolute SOH Estimation Error Using Conv-Net and Mobile-Net under Random Input SOC Ranges}
\label{fig:SOHEstimationErrorDistribution}
\end{figure}

Fig. \ref{fig:SOHEstimationErrorDistribution} shows the testing absolute error distributions for SOH estimation under random input SOC ranges satisfying constraints in Section \ref{Prep}. Results indicate that the proposed Mobile-Net has similar performance to Conv-Net for SOH estimation, but requires significantly lower computation and memory (to be discussed in Section \ref{OnboardImplementation}).

\begin{figure}[H]
    \centering
    \begin{subfigure}[h]{0.4\textwidth}
        \centering
        \includegraphics[width=\textwidth,trim=5 10 5 5,clip]{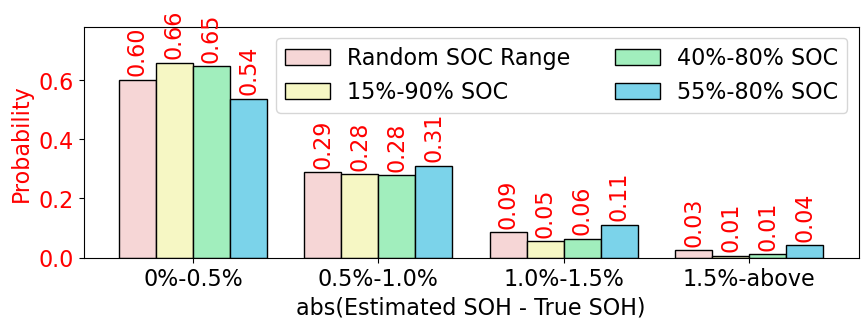}
        \caption{Conv-Net}
        \label{fig:Module_ConvNet_SOH_Estimation_DiffRange.png}
    \end{subfigure} \hfill

    \begin{subfigure}[h]{0.4\textwidth}
        \centering
        \includegraphics[width=\textwidth,trim=5 10 5 5,clip]{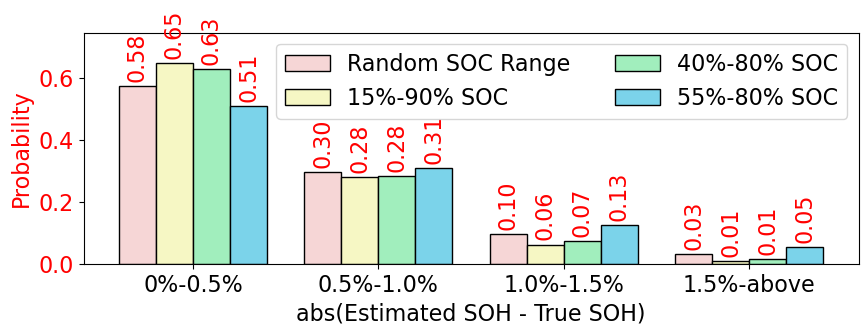}
        \caption{Mobile-Net}
        \label{fig:Module_MobileNet_SOH_Estimation_DiffRange}
    \end{subfigure} 
\caption{Testing Absolute Error Distributions for SOH Estimation under Different Input SOC Ranges}
\label{fig:Module_SOH_Estimation_Error_Distribution_Specific_Ranges}
\end{figure}

To assess the generalizability of the models trained from random SOC ranges to specific SOC ranges, the same three cases in Section \ref{CurvesResult} are employed. Similarly, the proposed Conv-Net and Mobile-Net are not re-trained, and all the data are used for testing. Fig. \ref{fig:Module_SOH_Estimation_Error_Distribution_Specific_Ranges} shows the absolute error distributions for the testing set, demonstrating that SOH estimation accuracy remains consistent across different SOC ranges, with a slight deterioration under narrower input SOC ranges due to reduced informational content.

In summary, direct SOH estimation using Conv-Net and Mobile-Net not only outperforms VIC/VDV-based SOH estimation using U-Net and Mobile U-Net, but also generalizes well across arbitrary SOC ranges that satisfy constraints given in Section \ref{Prep}.

\subsection{Computational Footprint Comparison}
\label{OnboardImplementation}
The computation and memory demands are critical factors for neural networks. Their footprint can be measured by the total number of CNN parameters and floating-point operations. Table \ref{table:ParamNum} summarizes these quantities for the proposed CNNs.

Based on Table \ref{table:ParamNum}, one can have the following observations. First, most of the parameters in the Conv-Net and Mobile-Net are inherited from the contraction path of the U-Net and Mobile U-Net for extracting degradation-related features. Second, when comparing U-Net to Mobile U-Net and Conv-Net to Mobile-Net, the mobile CNNs contain approximately 1/3 of the parameters and the floating-point operations. As seen in Sections \ref{CurvesResult}, \ref{FeaturesResult}, and \ref{SOHResult}, Mobile U-Net and Mobile-Net maintain strong performance in VIC/VDV curve construction and SOH estimation, respectively, despite their reduced complexity.

\begin{table}[H]
\centering
\caption{Number of Parameters and Floating-Point Operations for Proposed CNNs}
\label{table:ParamNum}
\begin{tabular}{|c|c|c|} 
\hline
\textbf{Models} & \begin{tabular}[c]{@{}c@{}} \textbf{No. of Parameters}  \\ \textbf{(Total, Trainable, Fixed)} \end{tabular} & \begin{tabular}[c]{@{}c@{}} \textbf{No. of Floating-Point} \\ \textbf{Operations} \end{tabular} \\ 
\hline\hline
U-Net & (95503, 94323, 1180) & 5537954 \\ 
\hline
Mobile U-Net & (32425, 31385, 1040) & 1879739 \\ 
\hline
Conv-Net & (73361, 12991, 60370) & 2809101 \\ 
\hline
Mobile-Net & (27118, 4946, 22172) & 981243 \\ 
\hline
\end{tabular}
\end{table}

\subsection{Adaptivity to New Charging Protocols and Datasets}
\label{GeneralizabilityToNewConditions}
The neural network-based approach inherently has poor generalizability to previously unseen conditions \cite{CNN_Book}. However, this limitation can be effectively mitigated through transfer learning, which enables the trained CNN models to adapt to unseen dynamic charging profiles and new battery datasets. Only a small subset of early and late layers of the trained models is fine-tuned, while all remaining layers are kept fixed. The extent of fine-tuning can be adjusted depending on application requirements, with fine-tuning more layers generally leading to improved performance.

To demonstrate this capability, a public dataset \cite{Attia}, comprising lithium iron phosphate (LFP) cells, is utilized. The cells are charged using 224 distinct six-step multi-stage fast-charging protocols, none of which are encountered during the training of the CNN models presented in Sections \ref{CurvesResult}, \ref{FeaturesResult}, \ref{SOHResult}, and \ref{OnboardImplementation}. Basic information about the dataset is summarized in Table \ref{table:UnseenDatasets}, while detailed descriptions can be found in \cite{Attia}. The dataset is then randomly partitioned into 60\% training, 20\% validation, and 20\% testing sets.

\begin{table}[H]
\centering
\caption{Key Attributes of the New Battery Dataset}
\label{table:UnseenDatasets}
\begin{tabular}{|l|l|} 
\hline
\textbf{Attributes} & \textbf{Descriptions} \\ \hline\hline
Chemistry & LFP \\ \hline
Cell Nominal Capacity & 1.1Ah \\ \hline
No. of Cells & 233 \\ \hline
Cell-Level SOH Range & 100\% - 80\% \\ \hline
No. of Fast Charging Protocols & 224 \\ \hline
No. of Input-Output Pairs & 56320 \\ \hline
\end{tabular}
\end{table}

To adapt to these new charging protocols, transfer learning is applied. Note that, as mentioned previously, the extent of fine-tuning is not unique and depends on one's needs. For demonstration purposes, this paper adopts the following fine-tuning strategy. For all four trained CNN models (e.g., U-Net, Mobile U-Net, Conv-Net, and Mobile-Net), their first four convolutions or depthwise separable convolutions in the contraction path are fine-tuned. For VIC/VDV curve construction, in the expansion path, the last five convolutions plus a corresponding transposed convolution are fine-tuned for the U-Net, while the last five depthwise separable convolutions plus a corresponding upsampling are fine-tuned for the Mobile U-Net. Similarly, for direct SOH estimation, the last two convolutions or depthwise separable convolutions are fine-tuned for Conv-Net or Mobile-Net, respectively.

The fine-tuned U-Net and Mobile U-Net accurately construct VIC/VDV curves under new fast-charging profiles. Then, SOH estimation is conducted using the VIC peak height as the sole input feature \cite{Weng1} and its performance is summarized in Table \ref{table:New_Estimation_Result_Table}. Similarly, the fine-tuned Conv-Net and Mobile-Net achieve accurate SOH estimation under these previously unseen fast-charging profiles, with performance metrics reported in Table \ref{table:New_Estimation_Result_Table}. Overall, Table \ref{table:New_Estimation_Result_Table} demonstrates that, with transfer learning, the proposed CNN architectures exhibit strong adaptability to both unseen charging protocols and new battery datasets.

\begin{table}[H]
\centering
\caption{SOH Estimation Performance under Previously Unseen Fast-Charging Protocols Based on Testing Set}
\label{table:New_Estimation_Result_Table}
\begin{tabular}{|c|c|c|c|} 
\hline
\begin{tabular}[c]{@{}c@{}}\textbf{Neural}  \\ \textbf{Network} \end{tabular} & \textbf{Approach} & \textbf{RMSE} & \begin{tabular}[c]{@{}c@{}} \textbf{99.7th-Percentile} \\ \textbf{Absolute Error} \end{tabular} \\ 
\hline\hline
U-Net & \begin{tabular}[c]{@{}c@{}} Through VIC \\ Peak Height \end{tabular} & \begin{tabular}[c]{@{}c@{}} 0.90\% \\ SOH \end{tabular} & 3.10\% SOH \\ 
\hline
Mobile U-Net & \begin{tabular}[c]{@{}c@{}} Through VIC \\ Peak Height \end{tabular} & \begin{tabular}[c]{@{}c@{}} 0.90\% \\ SOH \end{tabular} & 3.17\% SOH \\ 
\hline
Conv-Net & \begin{tabular}[c]{@{}c@{}} Direct \\ Estimation \end{tabular} & \begin{tabular}[c]{@{}c@{}} 0.23\% \\ SOH \end{tabular} & 1.04\% SOH \\ 
\hline
Mobile-Net & \begin{tabular}[c]{@{}c@{}} Direct \\ Estimation \end{tabular} & \begin{tabular}[c]{@{}c@{}} 0.29\% \\ SOH \end{tabular} & 1.36\% SOH \\ 
\hline
\end{tabular}
\end{table}


\section{CONCLUSIONS}
\label{Conclusion}
This paper proposes a unified approach, shown in Algorithms \ref{alg:Method1} and \ref{alg:Method2}, that enables ICA/DVA-based degradation monitoring under dynamic charging profiles for any SOC ranges that satisfy Constraint (\ref{eqn:SOCwindow}) for the first time. 

By introducing the concepts of VIC/VDV curves, the proposed U-Net constructs VIC/VDV curves that accurately approximate actual IC/DV curves. A case study on module-level SOH estimation demonstrates the practical utility of VIC/VDV curves for degradation monitoring. Applied to a large experimental dataset of NMC622 modules with three parallel-connected cells, VIC/VDV features achieve an RMSE of 0.73\% and a 99.7th-percentile absolute error of 2.54\% for module-level SOH estimation. If SOH is the only information of interest, the proposed Conv-Net provides accurate SOH estimates directly. On the same dataset, Conv-Net achieves an RMSE of 0.64\% SOH and a 99.7th-percentile absolute error of 2.28\% SOH for module-level SOH estimation, while maintaining a smaller computational footprint compared to other CNNs in the literature. 

The proposed Mobile U-Net and Mobile-Net can replace U-Net and Conv-Net, respectively, to significantly reduce computation and memory requirements. These mobile CNNs reduce the numbers of parameters and floating-point operations to 1/3 of their counterparts, without losing performance. Specifically, for module-level SOH estimation, VIC/VDV features from Mobile U-Net achieve an RMSE of 0.79\% SOH and a 99.7th-percentile absolute error of 2.86\% SOH, while Mobile-Net achieves an RMSE of 0.68\% SOH and a 99.7th-percentile absolute error of 2.44\%. 

The promising results also lead to some directions for future work. First, the performance of the proposed method under SOH values below 80\% is worth investigating for applications beyond electrified vehicles. Second, because of the data-driven nature of the proposed CNNs, it is essential to further assess the performance and generalizability of the proposed methods under practical considerations for onboard implementation, including computational limitations of in-vehicle hardware and the efficient scalability of the framework to a large number of modules within a battery pack. Third, the performance of the proposed framework for more general charging/discharging conditions and other battery chemistries besides NMC622 and LFP will be investigated.


\bibliographystyle{Bibliography/IEEEtranTIE}
\bibliography{Bibliography/IEEEabrv,Bibliography/BIB_xx-TIE-xxxx}\ 

\end{document}